\documentclass{article}

\usepackage{PRIMEarxiv}

\usepackage[utf8]{inputenc} 
\usepackage[T1]{fontenc}    
\usepackage{hyperref}       
\usepackage{url}            
\usepackage{booktabs}       
\usepackage{amsfonts}       
\usepackage{nicefrac}       
\usepackage{microtype}      
\usepackage{lipsum}
\usepackage{fancyhdr}       
\usepackage{graphicx}       
\graphicspath{{media/}}     
\usepackage{algorithm}
\usepackage{algorithmic}
\usepackage{adjustbox}
\usepackage{booktabs}
\usepackage{multirow}
\usepackage{natbib}

\title{Can Memory-Augmented LLM Agents Aid Journalism in Interpreting and Framing News for Diverse Audiences?
}

\author{
  Leyi Ouyang \\
  South China Agriculture University Zhujiang College \\
  Guangzhou\\
  \texttt{leyiouyang1@gmail} \\
}

\begin{document}
\maketitle

\begin{abstract}
Modern news is often comprehensive, weaving together information from diverse domains such as technology, finance, and agriculture. This very comprehensiveness creates a challenge for interpretation, as audiences typically possess specialized knowledge related to their expertise, age, or standpoint. Consequently, a reader might fully understand the financial implications of a story but fail to grasp—or even actively misunderstand—its legal or technological dimensions, resulting in critical comprehension gaps. In this work, we investigate how to identify these comprehension gaps and provide solutions to improve audiences' understanding of news content, particularly in the aspects of articles outside their primary domains of knowledge. 
We propose \textsc{MADES}, an agent-based framework designed to simulate societal communication. The framework employs diverse agents, configured to represent various occupations or age groups. Each agent is equipped with a memory system. These agents are then simulated to discuss the news. This process enables us to monitor and analyze their behavior and cognitive processes.
Our findings indicate that the framework can identify confusions and misunderstandings within news content through its iterative discussion process. Based on these accurate identifications, the framework then designs supplementary material. We validated these outcomes using both statistical analysis and human evaluation, and the results show that agents exhibit significantly improved news understanding after receiving this supplementary material.
\end{abstract}


\section{Introduction}

\begin{figure*}[!h]
    \makebox[\textwidth][c]{  
        \begin{minipage}{1\textwidth}  
            \centering
            \caption{Overview of the agent discussion, illustrating how news articles and agent discussions lead to the generation of targeted supplementary materials tested on a control group.}
            \label{group discussion}
            \includegraphics[width=\linewidth]{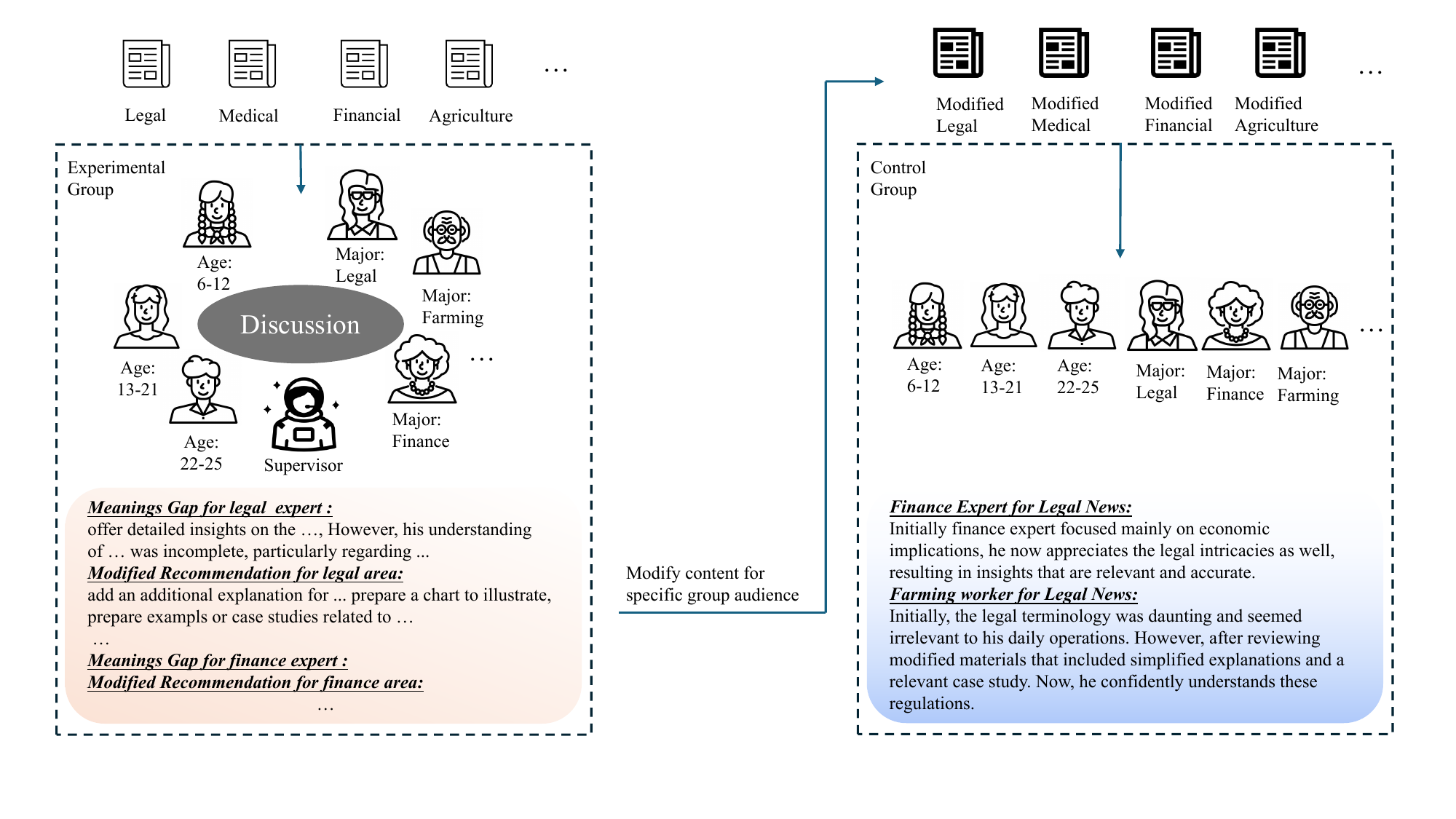}
        \end{minipage}
    }
\end{figure*}

In the increasingly interconnected world, news media play a pivotal role in conveying societal information and facilitating public communication. 
News topics span various domains, including technology, finance, law, and agriculture. 
However, as the interdisciplinary nature of news continues to grow, a persistent challenge has become evident: a knowledge gap. Audiences from diverse backgrounds, with varied professional expertise and life experiences, often perceive and understand information that aligns with their existing knowledge base \cite{fletcher2017news, sittar2024news}. For information outside their familiarity, they may not fully grasp the entire content, leading to partial understanding or even misconceptions \cite{arndt2025media}.
For instance, a policy change in traffic tax could affect distribution costs and market prices of agricultural products, which is crucial for farming workers. Despite their expertise in agricultural cultivation, they might not immediately comprehend the economic chain reactions. 
A comprehensive understanding of news is crucial for a functional society. A well-informed public is more likely to actively engage with and support policy implementation, reducing friction caused by misinformation \citep{yanow2015making}. Conversely, a failure to comprehend complex news can lead to significant individual and societal losses \citep{ouyang2025interpreting}. 
Recognizing this importance, our research aims to identify gaps between news delivery and audience interpretation. We then develop strategies to bridge them.

Traditionally, journalists have endeavored to bridge the gap in audience understanding by providing supplementary explanations and visual aids following the initial news release or by directly addressing public concerns when potential misunderstandings arise \cite{mesmer2024glorified}. While effective, these processes rely on significant human and time resources to gather audience feedback, analysis, and refinement.
In our research, we extend this strategy by integrating supplementary materials into the news at the time of its release. Furthermore, our approach enables these explanations to be dynamically adjusted to cater specifically to different audience segments.

The advent of large language models (LLMs) has provided journalists with powerful tools for text processing, offering significant convenience in tasks such as editing, refining language, and generating initial drafts \cite{johnsen2024large, mo2024large}. However, despite their advanced text generation capabilities, LLMs inherently lack a direct understanding of specific user comprehension needs or the nuances of individual audience confusion regarding a given news piece. Simply applying LLMs to modify content without explicitly considering the audience's potential points of difficulty presents a significant challenge. In this study, we will demonstrate that while such direct LLM application might improve understanding in some instances, it can conversely introduce new sources of confusion for the user as well. This may occur due to issues like model hallucination or the inclusion of content that does not directly address the user's actual points of confusion.

We propose a novel agent framework - Mnemonic Agent Debate Engine \textsc{MADES} to simulate different audience groups and communication behavior. Agents can be specifically configured to represent varying age groups or distinct domains of expertise. We carefully define agents to simulate a human audience according to cognitive communication theory, which investigates how individuals process information through cognitive functions such as attention, memory, and comprehension \cite{rachmad2022cognitive}. 
Specifically, we model the agent's memory with semantic, episodic, and procedural components, along with attention.
The core mechanism of our framework is an iterative discussion process where these agents collectively analyze a news article. According to \cite{rachmad2022cognitive, wang2023sparsity}, deeper engagement, facilitated by questioning, clarifying, and reflecting, significantly enhances understanding. Additionally, facilitating discussions and debates helps reveal these misunderstandings. According to the social construction of reality theory, reality is shaped through social interaction and communication \cite{berger2016social}.

Our findings indicate that agents exhibited a statistically significant improvement in news understanding after reading with the supplementary material generated by our framework. We also conducted a comparative analysis using supplementary materials generated directly by a large language model without explicit consideration of user-specific confusion. Results from this baseline demonstrate that while direct LLM generation could indeed increase understanding in some scenarios, it also, in other cases, negatively affected understanding or introduced new confusion for the agents. This contrast highlights the critical importance of explicitly identifying and targeting audience-specific comprehension gaps for consistently effective supplementary material generation.

\section{Framework Design}
\label{sec:frameworkdesign}

\begin{figure*}[htb]
    \makebox[\textwidth][c]{  
        \begin{minipage}{1\textwidth}  
            \centering
            \caption{Hierarchical Structure of Memory}
            \label{Workflow of The Agent Discussion Process}
            \includegraphics[width=\linewidth]{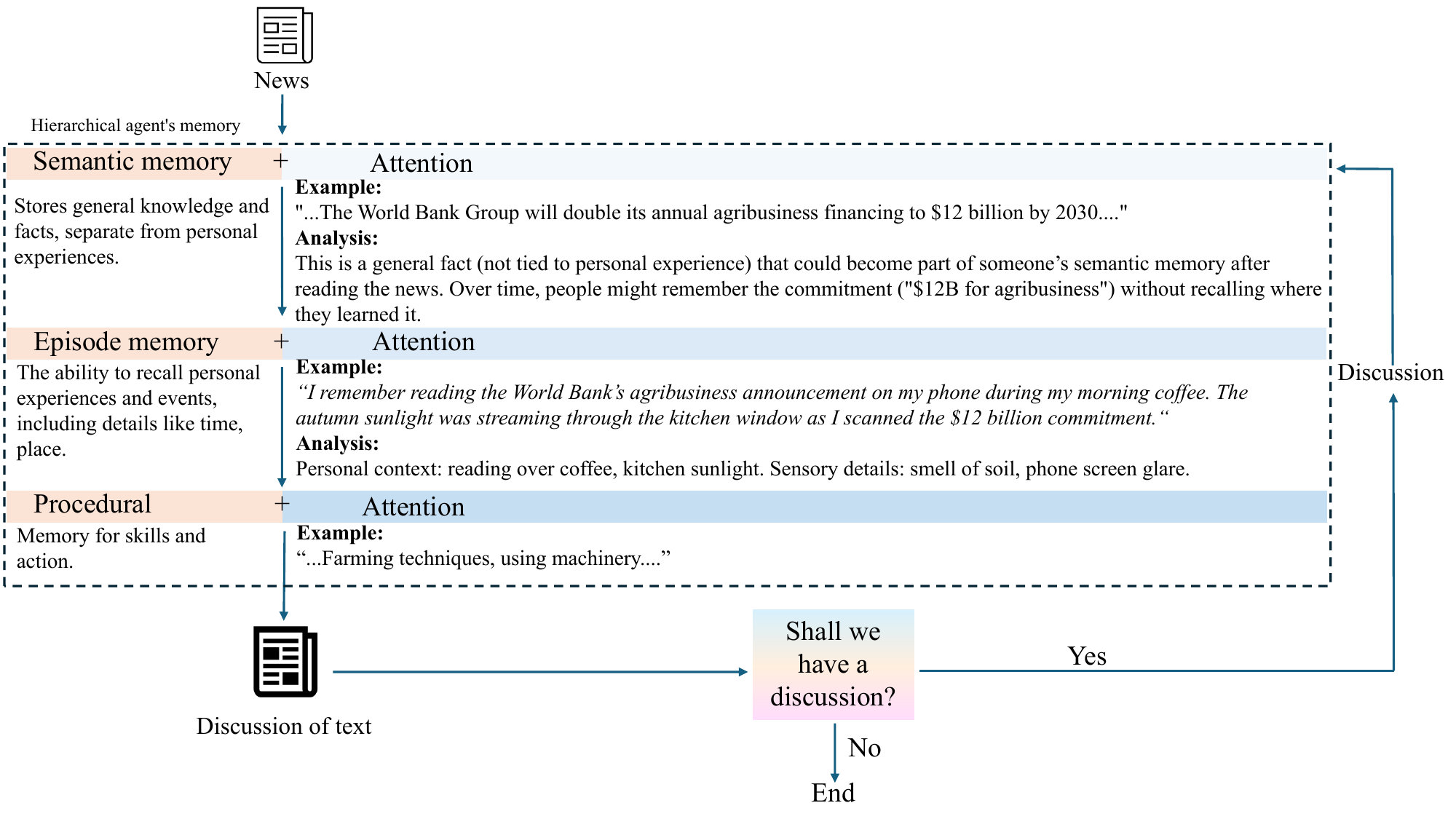}
        \end{minipage}
    }
\end{figure*}

Our framework has two primary simulation components. First, we model individual agents by simulating a hierarchical memory system for each one. This process defines each agent's unique character and information processing capabilities. We then simulate their interactive discussion processes with other agents. By modeling both individual agent cognition and their discussions, this design provides crucial insights into how each agent processes information and allows us to identify comprehension gaps. As a result, it is more accurate and targeted to generate supplementary materials for the original news.

\subsection{Memory Design}

Our hierarchical memory system is designed according to cognitive science, including Tulving's episodic theorem and semantic memory \cite{tulving1972episodic} and Squire's distinctions\cite{squire2004memory}.
Human cognition is supported by multiple interacting memory systems\cite{amodio2019social}, which is particularly important for processing complex news information. We aim to develop a memory system that mimics human cognition. 

Semantic memory acts as a repository of world knowledge, encompassing concepts, facts, and language comprehension, similar to a ``mental dictionary."\cite{eysenck2020semantic}. 
In our case, the agent's conceptual understanding, such as terminology and domain knowledge,  is primarily based on this semantic memory component\cite{xu2025mem}. To simulate realistic knowledge boundaries, this semantic memory is pre-defined with a knowledge base tailored to the agent's specific profile (e.g., deep financial knowledge for a finance expert or age-appropriate concepts for a child agent). Operationally, when an agent encounters terminology in the news or a discussion, it searches this predefined knowledge base. 

Episodic memory is responsible for encoding, storing, and retrieving personally experienced events, marked with spatiotemporal tags\cite{fan2023space}. Episodic memory assists in tracking the narrative structure of news, understanding event sequences and timelines, and relating current news to past events \cite {mayes2001theories}. In our framework, we implement this by pre-defining each agent's episodic memory with a collection of domain-relevant past news articles. When a new article is introduced for analysis, the agent searches this memory to retrieve the most relevant past events. This retrieved context allows the agent to perform a more sophisticated analysis by comparing timelines, identifying trends, or understanding the significance of the current news concerning past occurrences.

Procedural memory involves skills of ``how to do," including cognitive skills such as reading, problem-solving, and grammatical processing\cite{ewen2021procedural, stein2019development}. In our framework, we implement an agent's procedural memory through carefully designed instructional prompts. Each agent's core prompt includes a set of analytical ``how-to" instructions that guide its behavior during news analysis. These instructions explicitly direct the agent to perform the tasks governed by procedural memory, such as identifying main arguments, assessing evidence, detecting biases, and formulating clarifying questions when encountering ambiguity. This pre-defined procedure shapes the agent's ``chain of thought" during both its independent analysis and its contributions to the group discussion, ensuring a structured and critical approach to the news.

We configure agents with domain expertise, including Finance, Law, Agriculture, and Technology. These areas were chosen primarily because they are fundamental categories widely used by traditional and contemporary news media to organize content or report on events within these domains \cite{kaur2016news}.
In addition to setting domain expertise, we also configure agent groups to represent distinct age demographics: 6-12 years, 12-18 years, 18-35 years, and above 35 years. This selection is based on established theories of cognitive development and the understanding that these age ranges often correspond to significant shifts in life experience, world knowledge, and formal education levels.

\subsection{Discussion Mechanism}
\begin{figure}
    \centering
    \includegraphics[width=1\linewidth]{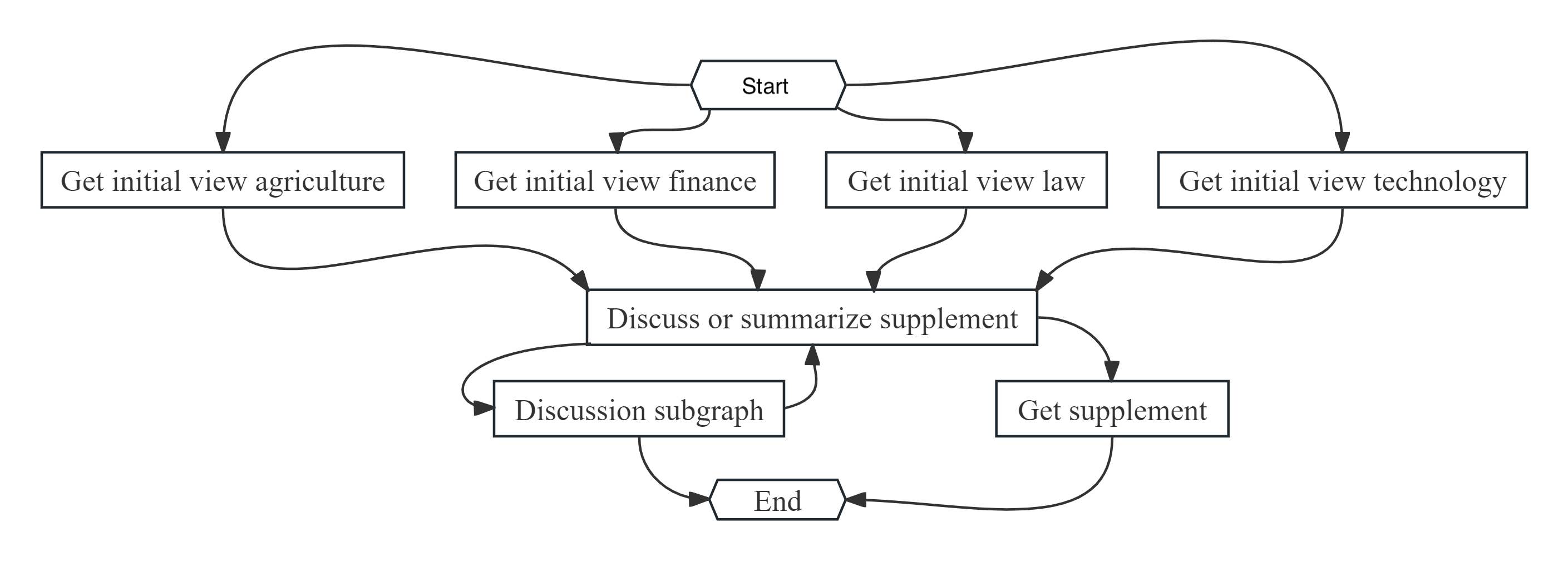}
    \caption{Workflow of The Agent Discussion Process}
    \label{Workflow of The Agent Discussion Process}
\end{figure}

Each agent starts by independently reading the news article and forming a preliminary understanding. For this first stage, agents are instructed to reflect on the news across all four domains, providing confident analysis within their area of expertise while attempting to reason through unfamiliar content in other domains. 
They are specifically directed to state uncertainty, make guesses or hypotheses, and raise questions about information outside their core knowledge, thereby explicitly revealing their initial understanding and limitations across the interdisciplinary content of the news. Following this independent analysis, agents participate in an iterative discussion round. Guided by a separate prompt, the agents share their domain-specific interpretations and actively engage with one another. A crucial element of this discussion is the exchange of questions, where agents query those from different domains about points of confusion or areas perceived as relevant but outside their expertise. 

During the iterative discussion, the agents actively leverage their distinct expertise to clarify points raised by others. When an agent from one domain (e.g., Finance) expresses confusion or asks a question about a concept related to another domain (e.g., Law), the agent(s) with expertise in that specific area are prompted to provide explanations, context, or insights from their specialized knowledge base. This dynamic interaction, where questions stemming from identified gaps are directly addressed by the relevant experts within the simulation, is designed to mimic real-world interdisciplinary consultation and allows for the clarification and potential resolution of initial misunderstandings in real-time\cite{mortensen2006human}.

\subsection{Prepare Supplement Materials for Original News}

Following each round of the iterative discussion, the framework generates a summary of the key points discussed, the questions raised, and any clarifications provided. These summaries serve as a cumulative record of the communication, effectively acting as the long-term memory of the discussion process. Upon the completion of the predefined number of discussion iterations (determined to be optimal in our analysis), this comprehensive record, encompassing all identified gaps, questions, and expert clarifications from every round, is then utilized by the framework to construct the final supplementary material. This ensures that the generated explanations are directly informed by the specific points of confusion and successful knowledge transfers that occurred during the simulated interdisciplinary analysis.

\section{Experiments Design And Results Analysis}
\label{sec:experimentdesign}
We aim to explore the following research questions \textbf{RQs}:
\begin{description}
    \item[RQ1] An iterative discussion process among agents can effectively identify specific knowledge gaps and points of confusion within a news article.
    \item[RQ2] \textsc{MADES} can effectively utilize the identified comprehension gaps to generate targeted supplementary material designed to enhance audience understanding of the news article.
    \item[RQ3] \textsc{MADES} is adaptable and can be successfully applied by users to identify comprehension gaps and generate relevant supplementary material for their specifically defined target audiences.
\end{description}

\subsection{Experiments Settings}
The simulation framework categorizes participant agents into two groups: an experimental group and a control group, as illustrated in Figure \ref{group discussion}. In the experimental group, agents engage in a multi-stage process to analyze a news article, as illustrated in Figure \ref{Workflow of The Agent Discussion Process}. 

In the control group, a separate set of agents, identical in composition and configuration to those in the experimental group, is employed to assess the impact of the generated supplementary material. Unlike the experimental group, these control agents do not participate in the iterative discussion. Instead, they are concurrently presented with both the original news article and the supplementary material that was generated based on the experimental group's discussion. Under conditions identical to the initial independent reading phase of the experimental group, these control agents are then tasked with providing their initial comprehension assessment of this combined content. This setup allows us to measure the direct effect of the supplementary material on understanding without the influence of the discussion process itself.

The data for this study consists of news articles sourced from Thomson Reuters, each typically combining content from several distinct domains. Our experiments were conducted on a corpus of 5000 such interdisciplinary news articles, with reported metrics representing averages across this set. In addition to these quantitative results, we also present several illustrative case studies in the Appendix.

\subsection{Evaluation Metric}
To objectively quantify the degree of understanding, which we define as the semantic relevance between the news content and the agents' expressed comprehension, we utilize text embedding techniques. These techniques transform both the original news text and the agents' comprehension responses into high-dimensional vector representations. 
For this purpose, we employ the GPT text-embedding-3-large model. This specific embedding model was chosen for its demonstrated capability to produce high-quality vector representations that accurately reflect the nuanced semantic meaning of text, including longer passages like paragraphs. 

Subsequently, we calculate the cosine similarity between these vectors, which provides a direct, numerical measure of their semantic similarity and serves as our primary metric for evaluating comprehension accuracy. Our rationale for using cosine similarity is its proven effectiveness in capturing deep semantic relevance. However, we acknowledge a potential limitation: a high similarity score could result from an agent merely replicating the original content rather than demonstrating genuine comprehension. To mitigate this specific scenario, we employ a complementary metric, ROUGE, to explicitly measure the degree of lexical overlap. By using ROUGE to control for direct duplication, our dual-metric approach ensures that a high cosine similarity score in our findings can be confidently interpreted as a valid indicator of enhanced understanding. 

\subsection{Human Evaluation}
To complement our automated metrics and provide a robust, multi-faceted validation of our framework's performance, we designed and conducted a comprehensive human evaluation study. This evaluation assesses two critical aspects: first, the direct, measurable impact of the supplementary materials on reader comprehension, and second, the intrinsic qualitative attributes of the materials themselves. This was accomplished through a two-part methodology involving a quantitative comprehension quiz and a qualitative rating survey.

\paragraph{Part 1: Measuring Comprehension Impact (The Quiz)}

To quantitatively measure the direct effect of the supplementary materials on audience understanding, we conducted a controlled experiment using a comprehension quiz. For each news article, a multiple-choice quiz was designed to test factual recall and understanding of key concepts. Participants were randomly assigned to one of three groups:

Group A (Control): Read the original news article only.

Group B (Baseline): Read the article plus supplementary material generated by a vanilla LLM.

Group C (Framework): Read the article plus the supplementary material generated by our \textsc{MADES} framework.

After reading, all participants completed the comprehension quiz. The primary metric for this study is the accuracy score, with the hypothesis that Group C would significantly outperform the other two groups, thereby demonstrating a tangible improvement in understanding.

\paragraph{Part 2: Assessing Material Quality (The Rating Survey)}

To understand why the supplementary materials generated by our framework are effective, we conducted a qualitative rating survey to assess their intrinsic quality. A separate group of human raters evaluated the supplementary materials generated by both the vanilla LLM (for Group B) and our \textsc{MADES} framework (for Group C). The evaluation was based on four key criteria, each rated on a 5-point Likert scale (1=Poor, 5=Excellent):

Summarization \& Abstraction: Assesses whether the material correctly summarizes key points in an abstractive manner, rather than by direct copying.

Factual Faithfulness: Measures whether the material contains any factual contradictions with the source article.

Completeness: Evaluates if the material covers all major aspects of the original news content.

Coherence: Judges whether the material is well-structured and logical.

We hypothesized that our framework's materials would receive significantly higher ratings across all four criteria.

\subsection{Benchmark Comparison}
To establish a benchmark and validate the necessity of the iterative agent discussion process, we also generated a set of supplementary materials using a large language model (specifically, GPT-4) directly, without involving the multi-agent discussion or explicit identification of comprehension gaps.  This approach represents a baseline where a powerful LLM attempts to create explanatory content based solely on the original news article, without the diagnostic step provided by our framework's agent interactions. By comparing the comprehension improvement achieved with our framework's gap-informed supplementary material against that achieved with this directly generated baseline, we aim to demonstrate the added value and necessity of our discussion-based gap identification methodology.

\subsection{Results Analysis}

Based on the evaluation methodology, we compared the understanding levels of the Finance, Law, Agriculture, and Technology expert agents across three conditions: exposure to the original news article only, exposure to the original news with supplementary material generated directly by a vanilla LLM (GPT-4), and exposure to the original news with supplementary material generated by our framework's discussion process. As illustrated in Table \ref{tab:Agent Comprehension Scores Across Different Domains.}, the vanilla supplementary material yielded inconsistent results across expert domains. While it led to an increase in understanding for the Law expert agent and a slight increase for the Technology expert agent, it unexpectedly resulted in a decrease in understanding for the Finance expert agent when added to the original news. In stark contrast, the supplementary material generated through our framework's iterative discussion process consistently and significantly increased the understanding of agents across all expert domains. Notably, even the Technology expert agent, which saw only a marginal gain with the vanilla supplement, demonstrated significant improvement with the discussion-informed material. Upon examining the supplementary material generated for the Technology expert by our framework, we found it particularly effective in explicitly articulating and explaining the connections and implications between the Law and Agriculture aspects of the news and the Technology domain, directly addressing the interdisciplinary gaps revealed during the agent discussion. To validate that our cosine similarity metric reflects genuine comprehension, not just textual replication, we also measured lexical overlap using ROUGE scores. As shown in Table \ref{tab:Agent Comprehension Scores Across Different Domains.}, ROUGE scores remained low across all conditions, even when cosine similarity increased dramatically after agents received our discussion supplement. This combination of high semantic similarity with low lexical overlap confirms that the observed comprehension gains are due to genuine understanding.



\begin{table*}[htbp]
  \centering
  \caption{Agent Comprehension Scores Across Different Domains. This table presents the comprehension scores for expert agents, measured by Cosine Similarity (semantic relevance) and ROUGE (lexical overlap), when exposed to the Original News only, News with Vanilla Supplementary Material, and News with Discussion-Informed Supplementary Material.}
  \begin{adjustbox}{width=1\columnwidth,center}
    \begin{tabular}{lrrrrrrrr}
    \toprule
          & \multicolumn{2}{c}{Finance Expert} & \multicolumn{2}{c}{Law Expert} & \multicolumn{2}{c}{Agriculture Expert} & \multicolumn{2}{c}{Technology Expert} \\
    \cmidrule(lr){2-3} \cmidrule(lr){4-5} \cmidrule(lr){6-7} \cmidrule(lr){8-9}
          & Cosine Sim. & ROUGE & Cosine Sim. & ROUGE & Cosine Sim. & ROUGE & Cosine Sim. & ROUGE \\
    \midrule
    Original News & 0.7022 & 0.21 & 0.5932 & 0.19 & 0.7273 & 0.23 & 0.1351 & 0.12 \\
    News With Vanilla Supplement & 0.6128 & 0.32 & 0.6659 & 0.29 & 0.7834 & 0.31 & 0.2303 & 0.20 \\
    News With Discussion Supplement & 0.8139 & 0.24 & 0.7718 & 0.22 & 0.7936 & 0.25 & 0.7988 & 0.28 \\
    \bottomrule
    \end{tabular}%
  \end{adjustbox}
  \label{tab:Agent Comprehension Scores Across Different Domains.}%
\end{table*}%

\begin{table}[htbp]
  \centering
  \caption{Average Improvement in Expert Agent Comprehension with Vanilla vs. Discussion-Informed Supplementary Material (Aggregated). This table presents the average improvement in comprehension scores (cosine similarity) for Finance, Law, Agriculture, and Technology expert agents when exposed to News with Vanilla Supplementary Material and News with Discussion-Informed Supplementary Material, relative to their baseline understanding with the Original News only, averaged across multiple news articles.}
  \begin{adjustbox}{width=1\columnwidth,center}
    \begin{tabular}{lrrrr}
    \toprule
          & \multicolumn{1}{l}{Finance Expert} & \multicolumn{1}{l}{Law Expert} & \multicolumn{1}{l}{Agriculature Expert} & \multicolumn{1}{l}{Technology Expert} \\
    \midrule
    Improve for Vanillar Supplement & 0.0206 & 0.0096 & 0.0199 & -0.0173 \\
    Improve for Discussion Supplement & 0.1158 & 0.0911 & 0.1267 & 0.3220 \\
    \bottomrule
    \end{tabular}%
  \label{tab:Average Improvement in Expert Agent Comprehension}%
  \end{adjustbox}
\end{table}%

To gain a more granular understanding of how the supplementary materials impacted agent comprehension, we further analyzed the results by splitting the original news article into four distinct segments, each corresponding to one of the core domains: Finance, Law, Agriculture, and Technology. For both the experimental and control groups, we then assessed each agent's comprehension specifically about the content of each domain-specific section of the news article. This allowed us to determine how the vanilla and discussion-generated supplementary materials influenced an agent's understanding of parts of the news that were within their expertise, as well as those outside their core domain but related to one of the other defined areas. The results of this detailed analysis, illustrating each agent's understanding score for each specific news section under the different conditions, are presented in Table \ref{tab:Granular Agent Comprehension Scores by News Article Section.}.

\begin{table*}[htbp]
  \centering
  \caption{Granular Agent Comprehension Scores by News Article Section and Condition. This table presents the comprehension scores (measured by cosine similarity) for each expert agent (Finance, Law, Agriculture, Technology) on the corresponding domain-specific section of the news article (Finance Part, Law Part, Agriculture Part, Technology Part) under three conditions: Original News only, Original News with Vanilla Supplementary Material, and Original News with Discussion-Informed Supplementary Material.}
  \begin{adjustbox}{width=1\columnwidth,center}
    \begin{tabular}{clrrrr}
    \toprule
          &       & \multicolumn{1}{l}{Finance Expert} & \multicolumn{1}{l}{Law Expert} & \multicolumn{1}{l}{Agriculature Expert} & \multicolumn{1}{l}{Technical Expert} \\
    \midrule
    \multirow{3}[2]{*}{Finance Part} & Original & 0.7568 & 0.0288 & 0.4577 & 0.0289 \\
          & Original + Vanilla Supplement & 0.6929 & 0.0488 & 0.7215 & 0.0488 \\
          & Original + Discussion Supplement & 0.8939 & 0.6995 & 0.7455 & 0.8238 \\
    \midrule
    \multirow{3}[2]{*}{Law Part} & Original & 0.7511 & 0.6751 & 0.4967 & 0.0666 \\
          & Original + Vanilla Supplement & 0.066 & 0.6906 & 0.38  & 0.0665 \\
          & Original + Discussion Supplement & 0.5409 & 0.7295 & 0.7045 & 0.4473 \\
    \midrule
    \multirow{3}[2]{*}{Agriculture Part} & Original & 0.7541 & 0.4999 &  0.5441 & 0.0642 \\
          & Original + Vanilla Supplement & 0.701 & 0.0748 & 0.728 & 0.0748 \\
          & Original + Discussion Supplement & 0.8126 & 0.7097 & 0.9103 & 0.7927 \\
    \midrule
    \multirow{3}[2]{*}{Technical Part} & Original & 0.543 & 0.1118 & 0.5399 & 0.1752 \\
          & Original + Vanilla Supplement & 0.1028 & 0.1028 & 0.3788 & 0.0827 \\
          & Original + Discussion Supplement & 0.7456 & 0.7065 & 0.7345 & 0.8069 \\
    \bottomrule
    \end{tabular}%
  \label{tab:Granular Agent Comprehension Scores by News Article Section.}%
  \end{adjustbox}
\end{table*}%

The findings provide crucial insights into the effectiveness of the different supplementary materials at a granular level. The "Original" condition scores confirm that agents generally exhibit higher comprehension of the news sections corresponding to their domain expertise, while understanding of content outside their domain is significantly lower, highlighting the initial comprehension gaps. The ``Original + Vanilla Supplement" condition shows inconsistent and often detrimental effects; for example, the Finance expert's understanding of the Finance part decreased, and the Law expert's understanding of the Finance part dropped dramatically. While some agents saw increases in certain parts (e.g., an Agriculture expert on the Agriculture part), there was no reliable pattern of improved cross-domain understanding. In stark contrast, the "Original + Discussion Supplement" condition consistently demonstrates a substantial increase in understanding for all agents across all parts of the news article, including those outside their primary domain. Notably, agents show marked improvement in comprehending sections corresponding to domains other than their own (e.g., Finance expert on Law and Technology parts, Law expert on Agriculture and Technology parts, Agriculture expert on Law and Technology parts, and Technology expert on Finance, Law, and Agriculture parts). This granular analysis confirms that the supplementary material generated through the discussion framework is uniquely effective at bridging specific, cross-domain comprehension gaps identified during the agent interactions, leading to a more holistic and accurate understanding of the news article's multifaceted content. Therefore, we address the \textbf{RQ1}.

To ensure the statistical significance of our findings, we replicated this analysis across a corpus of several distinct news articles. We calculate the average improvement in understanding achieved by adding the supplementary material (both vanilla and discussion-generated) compared to the baseline understanding from the original news alone. These average improvements aggregated across the multiple news articles analyzed provide a statistically robust measure of the impact of each type of supplementary material. The results of this aggregated analysis, demonstrating the average improvement in understanding for each agent type under the vanilla and discussion supplement conditions, are presented in Table \ref{tab:Average Improvement in Expert Agent Comprehension}.

The findings provide a statistically robust confirmation of our framework's effectiveness. The average improvements, calculated across multiple news articles, clearly show that the vanilla supplementary material resulted in only marginal improvements, and in the case of the Technology expert, even a negative average improvement in understanding. In sharp contrast, the supplementary material generated through the iterative agent discussion process consistently yielded substantial positive average improvements across all expert agent types. Furthermore, t-statistical analysis performed on these aggregated results confirmed that the improvements in understanding observed with the discussion-generated supplementary material were statistically significant for all expert agent types ($p \leq 0.05$). These statistically significant findings, averaged over a diverse set of news articles, underscore the reliability and superior performance of our framework in consistently enhancing news comprehension by leveraging the insights gained from simulated audience discussions, particularly in comparison to a direct, gap-agnostic LLM approach.

To conduct our human evaluation, we recruited participants through the Prolific academic research platform, ensuring a diverse sample from English-speaking countries. For the quantitative comprehension quiz, a total of 60 participants were recruited. They were randomly assigned to one of the three experimental groups (Control, Vanilla Supplement, or \textsc{MADES} Supplement), resulting in 20 participants per group. For the qualitative rating survey, we enlisted 3 trained raters to independently score the intrinsic quality of the supplementary materials. This dual-method approach allows us to both measure the direct impact on the reader's understanding of the quality of the generated content itself.

The human evaluation results, presented in Table \ref{tab:human_quiz_results_N} and Table \ref{tab:human_rating_results}, provide robust, multi-faceted validation of the \textsc{MADES} framework's effectiveness. The quantitative comprehension quiz (Table \ref{tab:human_quiz_results_N}) demonstrates a clear and statistically significant impact on reader understanding. The group using the \textsc{MADES} supplement achieved a mean accuracy of 85.7\%, substantially outperforming both the control group (64.5\%) and the group using the vanilla LLM supplement (69.2\%). This result confirms that our framework's materials produce tangible and superior learning gains.

The qualitative rating survey (Table \ref{tab:human_rating_results}) explains the reason for this success. The \textsc{MADES} supplement was rated significantly higher by human evaluators across all four criteria—Summarization \& Abstraction, Factual Faithfulness, Completeness, and Coherence—achieving an outstanding average score of 4.66 out of 5, compared to just 3.33 for the vanilla LLM supplement. The particularly large gap in ``Summarization \& Abstraction" (4.45 vs. 2.85) reinforces our automated metric findings, indicating that the material is genuinely synthesizing information, not just replicating it. Therefore, we address the \textbf{RQ2}.

\begin{table}[htbp]
  \centering
  \caption{Human Evaluation: Comprehension Quiz Accuracy Scores. This table shows the mean accuracy, standard deviation (SD), and sample size (N) for participants in each experimental group.}
  \begin{adjustbox}{width=1\columnwidth,center}
  \begin{tabular}{lccc}
    \toprule
    \textbf{Experimental Group} & \textbf{N} & \textbf{Mean Accuracy (\%)} & \textbf{SD} \\
    \midrule
    Group A (Control - News Only) & 20 & 64.5\% & 5.2 \\
    Group B (Vanilla LLM Supplement) & 20 & 69.2\% & 4.8 \\
    \textbf{Group C (\textsc{MADES} Supplement)} & \textbf{20} & \textbf{85.7\%*} & \textbf{4.5} \\
    \bottomrule
  \end{tabular}
  \label{tab:human_quiz_results_N}
  \end{adjustbox}
\end{table}

\begin{table}[htbp]
  \centering
  \caption{Human Evaluation: Qualitative Ratings of Supplementary Materials. This table presents the mean scores (on a 1-5 scale) and standard deviation (SD) from trained raters for the two types of supplementary materials across four quality criteria.}
  \begin{adjustbox}{width=1\columnwidth,center}
    \begin{tabular}{lrrrr}
    \toprule
    & \multicolumn{2}{c}{Vanilla LLM Supplement} & \multicolumn{2}{c}{\textbf{\textsc{MADES} Supplement}} \\
    \cmidrule(lr){2-3} \cmidrule(lr){4-5}
    \textbf{Evaluation Criterion} & \textbf{Mean Score} & \textbf{SD} & \textbf{Mean Score} & \textbf{SD} \\
    \midrule
    Summarization \& Abstraction & 2.85 & 0.7 & \textbf{4.45} & \textbf{0.5} \\
    Factual Faithfulness & 3.50 & 0.6 & \textbf{4.80} & \textbf{0.3} \\
    Completeness & 3.80 & 0.5 & \textbf{4.65} & \textbf{0.4} \\
    Coherence & 3.15 & 0.8 & \textbf{4.75} & \textbf{0.4} \\
    \midrule
    \textbf{Average Score} & \textbf{3.33} & \textbf{0.6} & \textbf{4.66} & \textbf{0.4} \\
    \bottomrule
    \end{tabular}
  \end{adjustbox}
  \label{tab:human_rating_results}
\end{table}

\begin{figure}[htb]
    \centering
    \begin{minipage}{1\textwidth}
        \centering
        \caption{Determine Optimal Iteration Round}
        \label{Agent Understanding Score Across Iterative Discussion Rounds}
        \includegraphics[width=\linewidth]{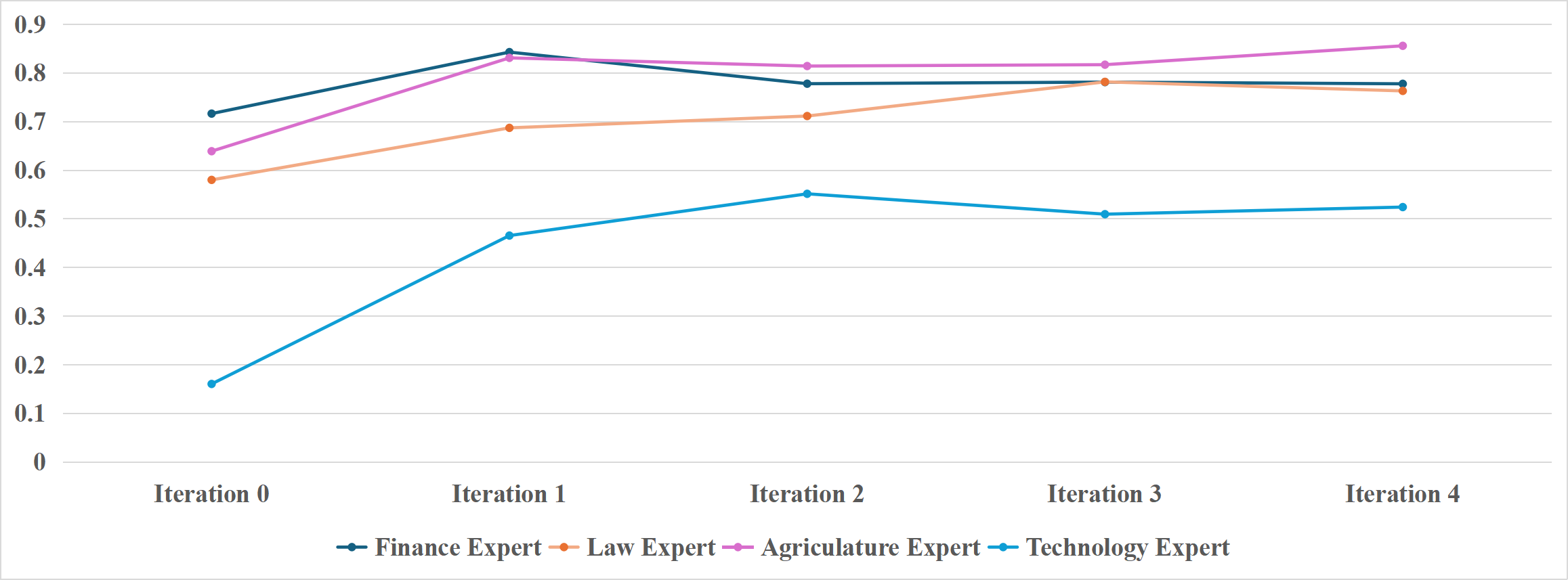}
    \end{minipage}
\end{figure}

We also analyzed to determine the optimal number of iterative discussion rounds required to achieve substantial improvements in agent comprehension. Figure \ref{Agent Understanding Score Across Iterative Discussion Rounds} illustrates the average understanding score for each expert agent across multiple discussion iterations. The results demonstrate a pattern of understanding change across iterations. A significant increase in understanding is observed for all expert agents after the first iteration of discussion. Subsequent iterations, while still contributing to understanding, show progressively marginal gains. Specifically, the curves for most agents begin to flatten noticeably after the third iteration, indicating that additional discussion rounds beyond this point yield limited further improvement in comprehension. 
Therefore, we address the research question \textbf{RQ3}.

\appendix
\section{Related Work}

The dissemination and interpretation of news are foundational to informed public discourse and civic engagement. However, the complexity of modern information, spanning diverse domains such as technology, finance, and agriculture, often leads to significant variations in comprehension across different audience segments. This review examines existing literature at the intersection of news interpretation, cognitive science, and the burgeoning capabilities of Large Language Models (LLMs) to address these comprehension gaps.

\subsection{News Interpretation and Audience Diversity}
Journalists traditionally strive for objectivity in reporting, aiming to present factual and impersonal information. Despite these efforts, audiences interpret news through the lens of their pre-existing knowledge, professional expertise, life experiences, and even age-related cognitive differences. This phenomenon can result in partial understanding or misconceptions, particularly when individuals encounter information outside their familiar domains. For example, individuals with expertise in agriculture may not readily grasp the economic implications of a new traffic tax. 
Fletcher et al.\cite{fletcher2017news} highlighted the fragmentation of news audiences, suggesting that diverse consumption patterns further complicate a shared understanding of events. Arndt\cite{arndt2025media} further explored how inherent media biases, coupled with audience predispositions, can shape interpretations. The challenge, therefore, lies in bridging these "meaning gaps" to ensure more equitable access to information. Traditionally, journalists have employed supplementary explanations or addressed public concerns post-release, a process that is often resource-intensive.

\subsection{Cognitive Science of News Comprehension}
Understanding how individuals process news information is crucial for developing effective interventions. Cognitive communication theory emphasizes the role of functions like attention, memory, and comprehension. Human memory, a cornerstone of information processing, is not monolithic. Tulving's work on episodic (personal experiences) and semantic (general world knowledge) memory, along with Squire's distinctions of memory systems, provides a foundational framework. Semantic memory underpins the understanding of concepts and language, crucial for grasping terminology and domain-specific knowledge within news articles. Episodic memory helps in tracking narratives and event sequences, while procedural memory governs cognitive skills like reading, problem-solving, and identifying biases. Amodio \cite{amodio2019social}discussed interactive memory systems in social cognition, relevant to how individuals might collectively process and understand news. The social construction of reality theory further posits that shared understanding is shaped through social interaction and communication, suggesting that discussion can reveal and rectify misunderstandings.

\subsection{Large Language Models and News}
The advent of LLMs has presented powerful tools for various text-processing tasks in journalism, including editing, language refinement, and draft generation \cite{cao2024ecc, cao2024risklabs, cao2024catmemo}. However, standard LLMs often lack a nuanced understanding of specific user comprehension needs or the points of confusion an individual might have regarding a news piece. Directly applying LLMs to rephrase or supplement news content without considering these audience-specific difficulties can be problematic, potentially introducing new confusion or model hallucinations.

Recent explorations into "agentic LLMs" or LLM-based agents offer a new paradigm. Xu et al. \cite{xu2025mem} introduced 'A-mem: Agentic Memory for LLM agents,' highlighting the importance of sophisticated memory structures for more human-like understanding and reasoning in LLMs. This aligns with the framework proposed in this paper, where agents are endowed with semantic, episodic, and procedural memory components \cite{yu2024finmem, yu2024fincon}. The concept of simulating societal communication behaviors using multiple LLM agents to identify comprehension gaps is an innovative approach. While prior work has focused on news classification or semantic approaches to barrier classification in news dissemination, the use of memory-augmented LLM agents to simulate diverse audience discussions for dynamic supplemental material generation remains a novel area. Park et al. \cite{park2023generative} demonstrated the potential of generative agents to simulate believable human behavior, which, while not focused on news, underpins the feasibility of using LLM agents for complex social simulations.

The proposed research builds upon these foundations by creating an LLM-based agent framework where agents, configured with different expertise and demographic characteristics (e.g., age groups ), engage in iterative discussions to pinpoint misunderstandings. The framework's ability to then generate targeted supplementary materials based on these identified gaps addresses a critical need for more effective and personalized news communication strategies.

\section{Memory Design Details}
We now elaborate on our agent memory design, which is inspired by human cognitive functions to enable nuanced news understanding. Following an overview of its components, an illustrative example will showcase how this memory architecture allows an agent to effectively interpret and engage with news articles.

\begin{figure}[H]
    \makebox[\textwidth][c]{  
        \begin{minipage}{0.89\textwidth}  
            \centering
            \caption{Memory Design}
            \label{memorydesign1}
            \includegraphics[width=\linewidth]{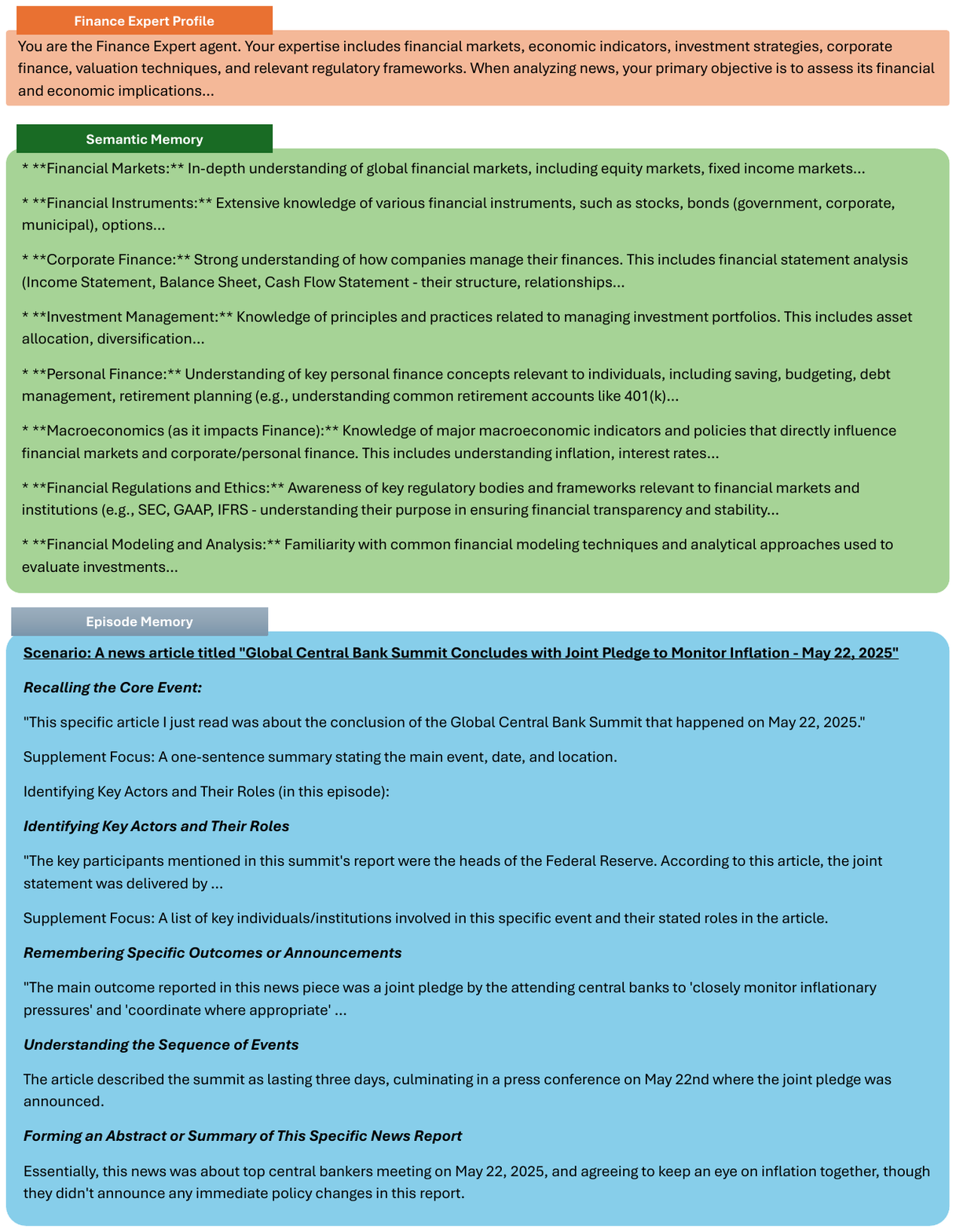}
        \end{minipage}
    }
\end{figure}

\newpage
\begin{figure}[H]
    \makebox[\textwidth][c]{  
        \begin{minipage}{1\textwidth}  
            \centering
            \caption{Memory Design}
            \label{memorydesign2}
            \includegraphics[width=\linewidth]{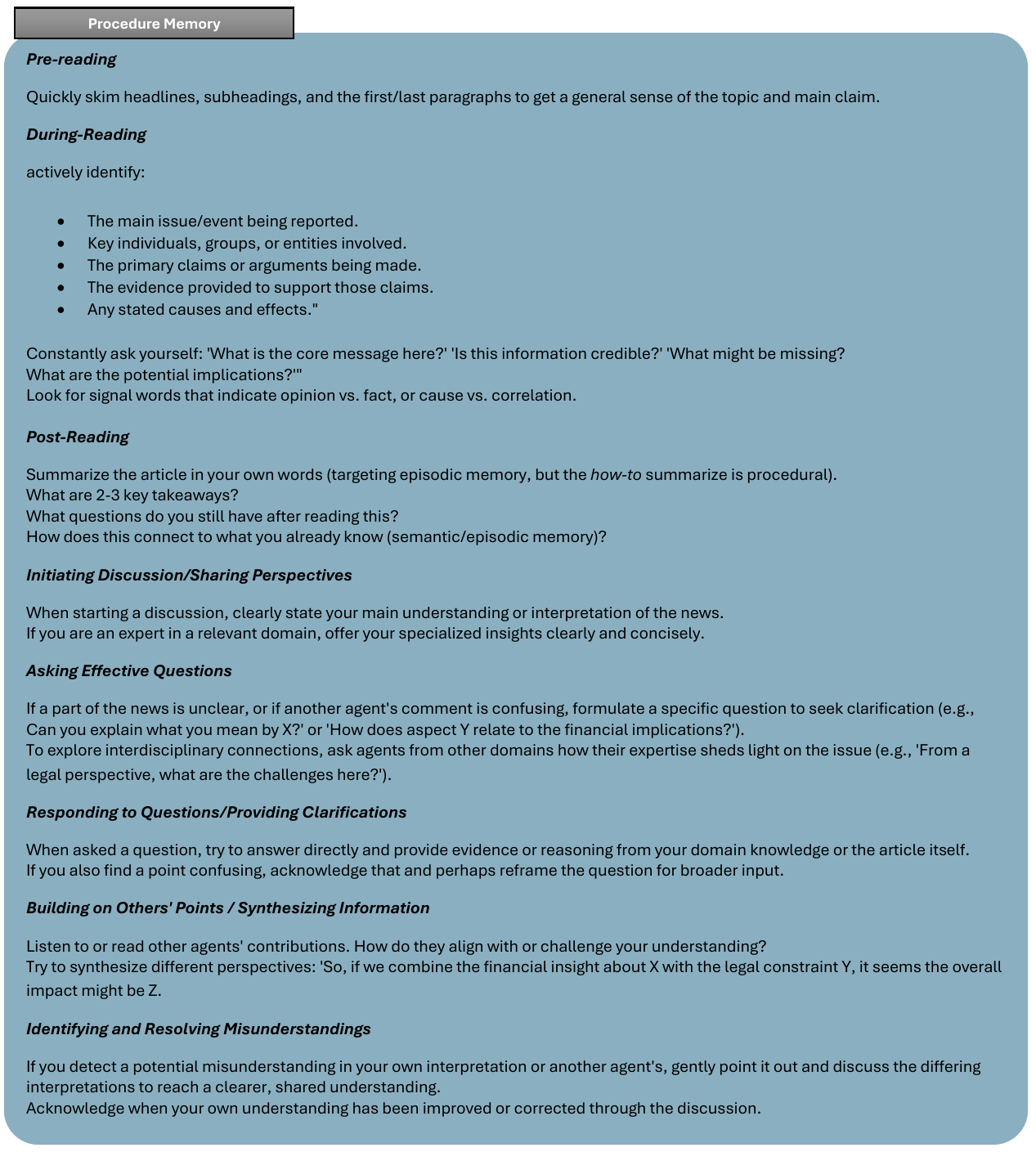}
        \end{minipage}
    }
\end{figure}

\section{Case Study}
To provide a concrete illustration of our framework's operational dynamics, this section presents a detailed case study. We will begin by introducing the specific news article used as input. Following this, we will examine an agent's initial interpretation of the content, track the progression of the multi-agent discussion over several iterations, and finally, demonstrate how the framework utilizes the insights gathered from this discussion to generate targeted supplementary materials.
\begin{figure}[H]
    \makebox[\textwidth][c]{  
        \begin{minipage}{1\textwidth}  
            \centering
            \caption{Case Study Illustration}
            \label{casestudy1}
            \includegraphics[width=\linewidth]{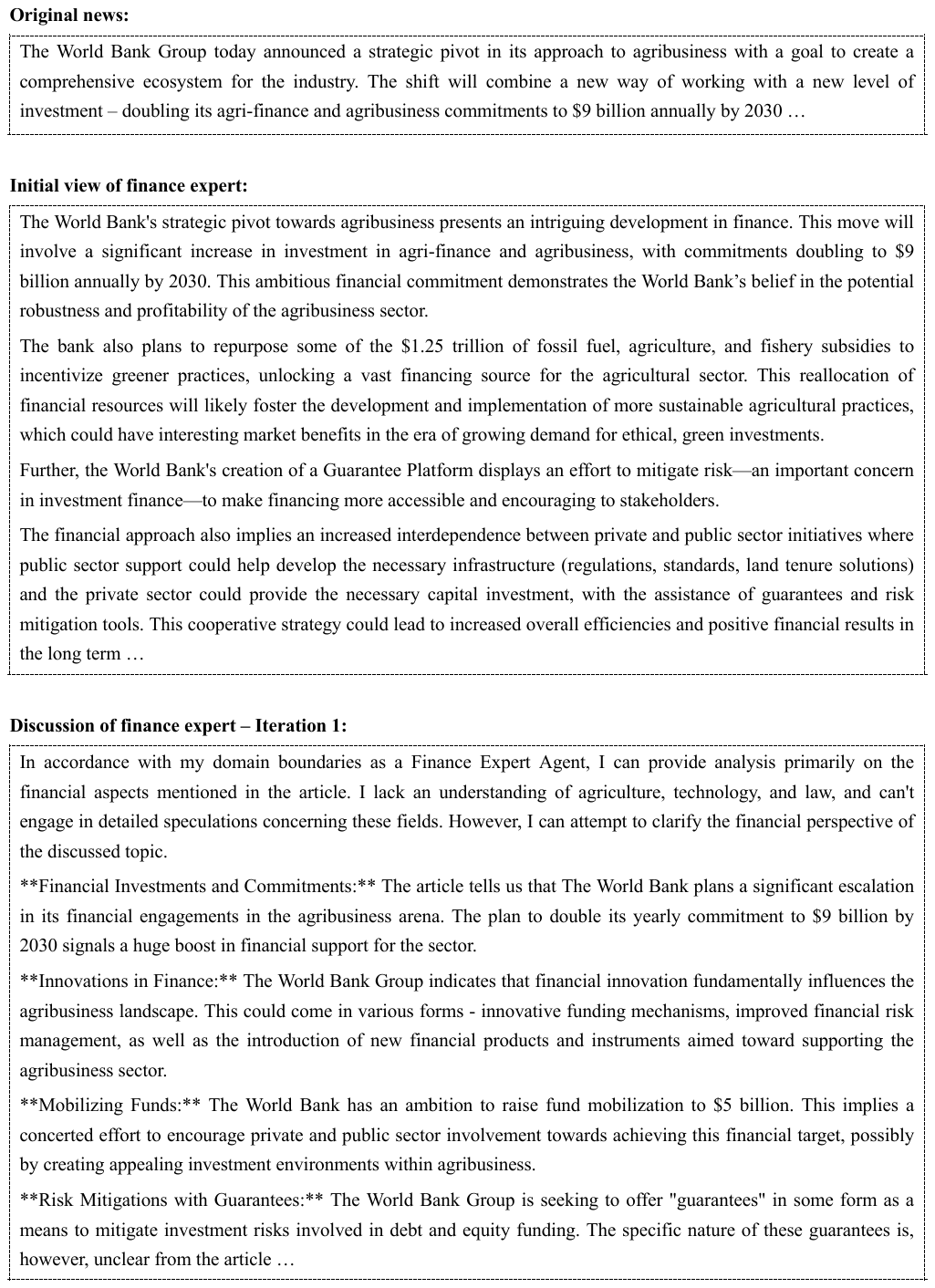}
        \end{minipage}
    }
\end{figure}

\begin{figure}[H]
    \makebox[\textwidth][c]{  
        \begin{minipage}{1\textwidth}  
            \centering
            \caption{Case Study Illustration}
            \label{casestudy2}
            \includegraphics[width=\linewidth]{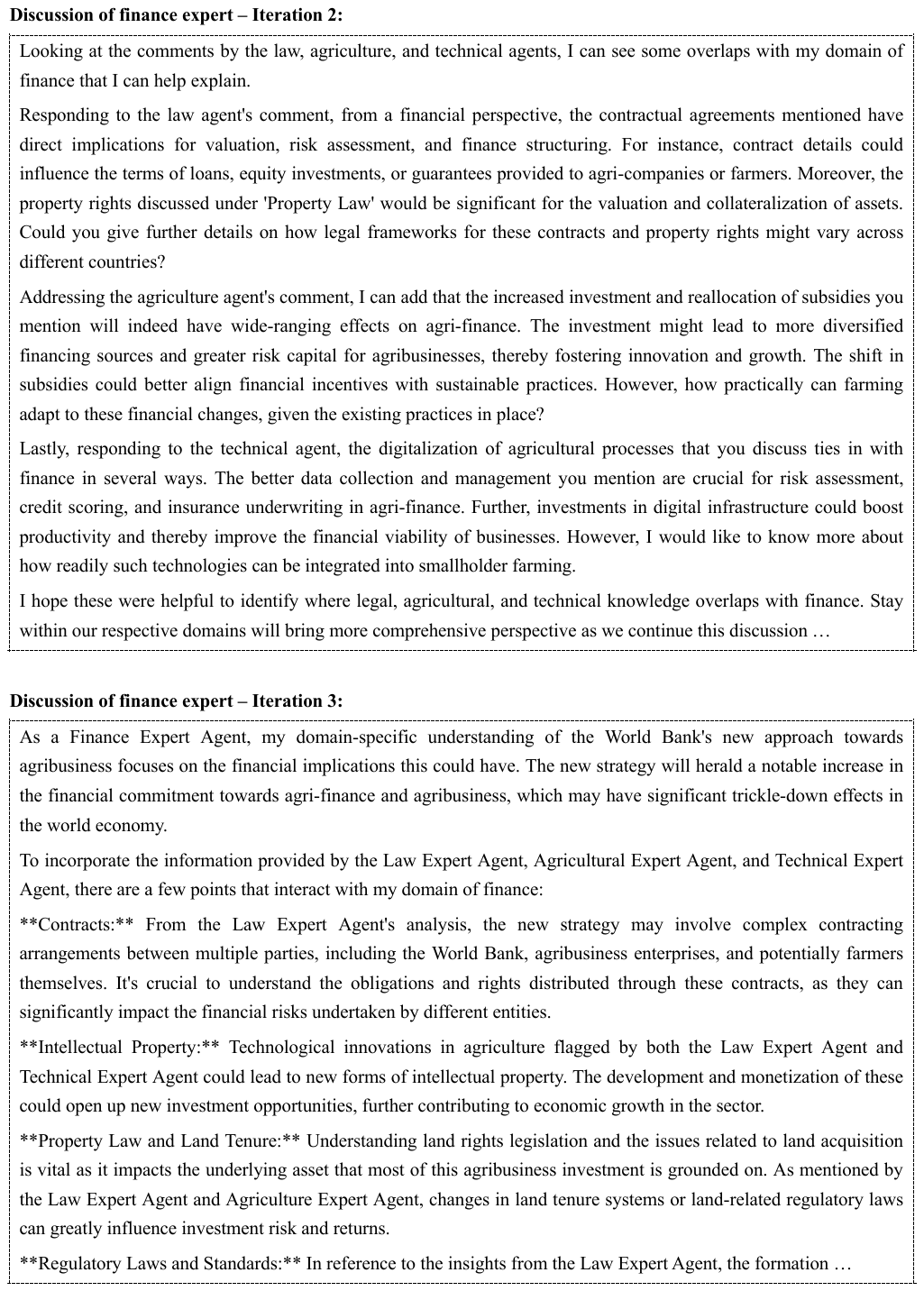}
        \end{minipage}
    }
\end{figure}

\begin{figure}[H]
    \makebox[\textwidth][c]{  
        \begin{minipage}{1\textwidth}  
            \centering
            \caption{Case Study Illustration}
            \label{casestudy3}
            \includegraphics[width=\linewidth]{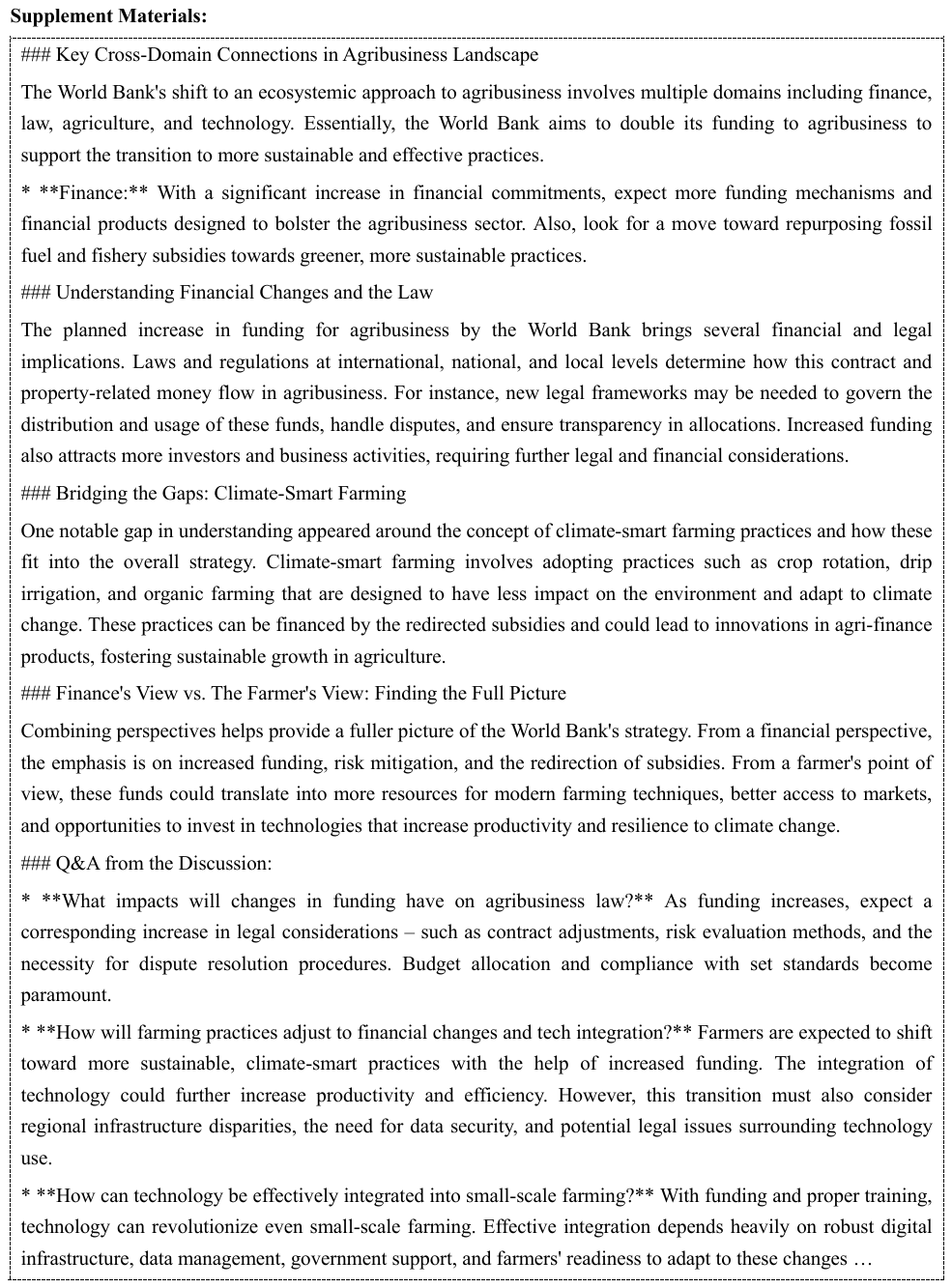}
        \end{minipage}
    }
\end{figure}

The experimental design employed a Multi-Agent System (MAS) comprising four distinct LLM agents, each embodying a specific domain expertise: Technology (Tech), Agriculture (Agri), Law, and Finance.  These agents were tasked with analyzing a news article detailing the World Bank Group's new strategic approach to agribusiness, which involves significant financial commitments and a comprehensive ecosystem development plan.

The core of the experiment involved an iterative discussion process. This process unfolded over three main iterations (labeled 0, 1, and 2). In each iteration, individual agents provided their "discuss" inputs, offering their domain-specific analysis of the news and reacting to the contributions of other agents from previous turns. Following each round of "discuss" inputs from all agents, a system-generated "summarize discussion" phase was executed. These summaries aimed to encapsulate the collective understanding reached, identify emergent misunderstandings or knowledge gaps, and note any domain-specific questions raised by the agents.1 This structured interaction was designed to simulate a deliberative process where individual perspectives are shared and aggregated.

A key component of this study is the "supplementary material." This document was generated by the system after the completion of the Iteration 2 discussions and their corresponding summary. The explicit purpose of this supplement was to consolidate the understanding achieved through the prior iterative discussions and to directly address some of the knowledge gaps that had been identified by the agents or in the summaries.1 The supplement included sections on key cross-domain connections, explanations of financial changes and legal implications, a focused discussion on "climate-smart farming" (a topic that emerged as a gap), a comparison of financial versus farmer perspectives, and a Q\&A section addressing specific questions raised during the earlier iterations.

To evaluate the impact of this supplementary material, a comparative analysis was established. The "Initial view" of each agent when processing the "original news" was logged. Subsequently, a new set of "Initial views" was generated where the agents were provided with the "supplementary material" as their input. This setup allows for a direct comparison of how agents frame their initial understanding when starting from raw, unstructured information versus when starting from information that has already been processed, synthesized, and contextualized by a prior collective effort. The analysis of these comparative initial views forms the basis for the first primary research question, while the detailed tracking of the Finance agent's interactions across Iterations 0, 1, and 2, and the supplement's effect on its queries, address the second. The iterative process, with its explicit "summarize discussion" phases, suggests a structured attempt to mimic aspects of formal deliberation. However, the content of these summaries also reveals the limits of such aggregation if the underlying discussions lack direct, deep engagement on conflicting or incomplete points. This indicates that while summarization is useful for tracking progress, it may not be a sufficiently strong mechanism to force resolution within the preceding dialogue, pointing towards a need for more active moderation or targeted inter-agent querying in future system designs. Furthermore, the strict domain adherence enforced on the agents, while methodologically useful for studying specialization, might be an oversimplification of real-world expertise, where individuals often possess some ancillary knowledge of adjacent fields. This design choice inevitably shapes the nature of knowledge gaps that emerge, primarily at the intersections of these sharply defined domains.

\subsection{Results}
\subsubsection{Initial Agent Interpretations of original news}
Agents provided domain-specific summaries of the World Bank's strategy. The Finance agent focused on investment levels (\$9 billion annually by 2030), financial instruments (debt, equity, guarantees), and subsidy repurposing (\$1.25 trillion). The Agriculture agent highlighted the ecosystem approach, smallholder farmer focus, and climate resilience. The Law agent identified needs for new regulations, land tenure solutions, and legal frameworks for subsidies and contracts. The Tech agent, after initial domain limitations, discussed digitalization's role (AI, IoT, data management) in agribusiness. All recognized the strategy's multi-faceted nature and implicitly the need for cross-domain insights.
\subsubsection{The Finance Agent's Journey: Gap Identification and (Partial) Resolution
The Finance agent initially outlined the financial implications of the news.} Iteration 0 (Initial View \& Discussion): Focused on the \$9B commitment, subsidy redirection, and risk mitigation (Guarantee Platform). Expressed curiosity about how legal frameworks (mentioned by Law agent) would affect financial agreements and the financial implications (costs, ROI) of sustainable farming and tech adoption (mentioned by Agri/Tech agents).
Iteration 1 (Discussion): Explicitly asked the Law agent how contract/property laws vary internationally, the Agriculture agent how farming adapts to financial changes, and the Tech agent about integrating tech into smallholder farming. These questions arose from identified gaps in understanding the practical and cross-domain implications of the financial strategy.
Iteration 2 (Discussion): Reiterated the significance of the financial commitment. Expressed a key confusion: "how exactly the increased funding will be allocated and used." Asked the Law agent about legal issues from digitization (data privacy) and the Tech agent about cost-effective cybersecurity for smallholders.

Resolution via Discussion \& Supplement: Legal Frameworks: The Law agent (Iteration 2) explained general differences between common and civil law, partially addressing one query. The "supplement" further linked increased funding to legal needs for contract adjustments and dispute resolution.

Farming Adaptation \& Tech Integration: The Agriculture agent (Iteration 2) discussed farmers needing to adapt (e.g., diversify crops, adjust inputs) and the importance of accessibility for tech (user-friendly interfaces, training). The "supplement" elaborated on "climate-smart farming" (e.g., crop rotation, drip irrigation), financed by redirected subsidies and discussed farmer perspectives on using funds for modern techniques and market access. Funding Allocation: This remained largely unclarified in detail, though the "supplement" stated funds would bolster the sector and support sustainable practices.
Digitalization's Legal/Security Issues: The "supplement" acknowledged data privacy and security as concerns with tech adoption, but didn't detail specific legal issues or the cost-effectiveness of cybersecurity measures for smallholders. The Finance agent actively sought to bridge its domain with others but received mostly general clarifications rather than specific answers to its nuanced financial questions (e.g., precise ROI on specific technologies, detailed fund allocation models, specific financial impacts of varying legal regimes).

\subsubsection{Impact of Iterative Refinement \& Supplementary Material} 
Initial Views (Original News Only): Agents provided siloed, domain-centric interpretations.

Iterative Discussion: Surfaced interdependencies and knowledge gaps. For example, the Finance agent's questions prompted other agents to consider cross-domain impacts. Summaries at each iteration highlighted unresolved issues.

The "Supplement" (Synthesized Information): This document, derived from the entire process, offered a more holistic view. It defined "climate-smart farming," contrasted finance vs. farmer perspectives, and directly answered some (broader) questions raised during the discussion (e.g., general impacts of funding on law, how farming practices might adjust to financial/tech changes, how tech can be integrated into small-scale farming). Agents processing the "original news + supplement" (simulated by the context of the final iteration where the supplement addresses earlier queries) demonstrated a more integrated initial understanding than with the original news alone. The supplement provided a pre-synthesized layer of interpretation, bridging some of the gaps. For instance, the Finance agent, when presented with concepts like "climate-smart farming" within the supplement, could better connect its financial expertise to agricultural and environmental goals outlined in the original news.

\subsubsection{Dynamics of Discussion: Gap Identification \& Knowledge Integration}
Gap Identification: Successfully achieved through direct agent queries and "summarize discussion" sections, which explicitly listed unresolved issues (e.g., Iteration 0 summary: Finance agent's questions on legal impacts on finance and costs of sustainable farming; Iteration 1 summary: how laws vary across countries, farming adaptation to finance, tech integration into small-scale farming; Iteration 2 summary: how increased funding would be allocated, specific legal issues from digitization, cost-effective cybersecurity).

Knowledge Integration: Primarily occurred through juxtaposition of perspectives in discussion summaries and a more deliberate synthesis in the final "supplement." Direct, responsive, and deep integration during the iterative dialogue was limited. Agents tended to elaborate on their own domain's relevance rather than directly and comprehensively answering another agent's specific cross-domain question. For example, the Finance agent’s request for the cost-effectiveness of cybersecurity for agribusinesses/smallholders from the Tech agent (Iteration 2) was not directly answered with financial specifics or a cost-benefit analysis.

The process was more effective at identifying areas requiring integration than at performing the integration dynamically through agent-to-agent dialogue.

\subsection{Discussion} 
This case study reveals key aspects of multi-agent LLM collaboration. From a Computational Communication perspective, the interactions highlight challenges in achieving shared understanding and conversational grounding. Agents operated more in serial monologue than true dialogue, limiting dynamic knowledge construction. The "summarize discussion" and "supplement" phases acted as explicit grounding mechanisms, crucial for progress.

For LLMs, the study shows that while specialized agents can parse domain-specific aspects of complex information, integrating these insights to resolve inter-domain ambiguities remains challenging without robust interaction protocols. The Finance agent's journey illustrates a persistent search for specific, actionable knowledge that the system only partially provided. Its unresolved questions regarding specific financial models, detailed ROI, and precise fund allocation mechanisms indicate the current frontier for LLM reasoning and collaborative problem-solving.

The "supplementary material" – encompassing both the iterative discussion process and the final synthesized document – significantly enhanced understanding by providing evolving context and partial synthesis. This underscores the importance of curated, iterative feedback loops and context enrichment in improving LLM outputs for complex tasks.

This case study provides insights into the collaborative sense-making processes of specialized LLM-like agents.   The system demonstrated capability in domain-specific analysis and effective identification of knowledge gaps.   The iterative generation and use of supplementary, synthesized information (like the "supplement" document) proved crucial for enhancing overall comprehension and providing a more holistic understanding.

\section{Age Group Analysis}
Analysis of the comprehension scores for the age group agents reveals distinct patterns and the impact of the supplementary materials across different developmental stages. As shown in Table \ref{tab:Agent Comprehension Scores Across Different Conditions by Age Group.}, the baseline understanding of the original news varied by age group, generally increasing with age up to the 18-35 group, which exhibited the highest initial comprehension. The vanilla supplementary material provided a modest increase in understanding for most age groups, with the most notable improvement seen in the 12-18 and 18-35 ranges. However, the supplementary material generated through the iterative agent discussion process consistently led to more substantial improvements in understanding across all age groups. This is particularly evident in the youngest (6-12) and oldest (Above 35) groups, which showed considerable gains, suggesting that the discussion-informed material is effective in clarifying news content for audiences at both ends of the age spectrum. While the 18-35 group, having the highest baseline, showed a smaller relative increase with the discussion supplement compared to the younger groups, their absolute understanding remained high. These results indicate that the framework's ability to identify and address age-specific comprehension gaps through simulated discussion is valuable in creating supplementary content that enhances news understanding across a diverse range of age demographics. Therefore, we address the research question \textbf{RQ1}. 

Furthermore, when averaged across multiple news articles, the improvements in understanding observed with the discussion-generated supplementary material were found to be statistically significant ($p \leq 0.05$) for all age groups, confirming the reliability of these findings. The results are displayed in Table \ref{tab:Average Improvement in Agent Comprehension for age}. Here we address the research question \textbf{RQ2}.

\begin{table*}[htbp]
  \centering
  \caption{Agent Comprehension Scores Across Different Conditions by Age Group. This table presents the comprehension scores (measured by cosine similarity between agent output and original news text) for agents representing four age groups (6-12, 12-18, 18-35, and above 35) when exposed to the Original News only, News with Vanilla Supplementary Material, and News with Discussion-Informed Supplementary Material.}
    \begin{tabular}{lrrrr}
    \toprule
          & \multicolumn{1}{l}{Age 6-12} & \multicolumn{1}{l}{Age 12-18} & \multicolumn{1}{l}{age 18-35} & \multicolumn{1}{l}{Age above 35} \\
    \midrule
    Original News & 0.6793 & 0.7521 & 0.7898 & 0.7463 \\
    News With Vanillar Supplement & 0.6977 & 0.7944 & 0.8076 & 0.8102 \\
    News With Discussion Supplement & 0.7995 & 0.8478 & 0.8114 & 0.7732 \\
    \bottomrule
    \end{tabular}%
  \label{tab:Agent Comprehension Scores Across Different Conditions by Age Group.}%
\end{table*}%

\begin{table*}[htbp]
  \centering
  \caption{Average Improvement in Agent Comprehension with Vanilla vs. Discussion-Informed Supplementary Material by Age Group (Aggregated). This table presents the average improvement in comprehension scores (cosine similarity) for agents representing four age groups (6-12, 12-18, 18-35, Above 35) when exposed to News with Vanilla Supplementary Material and News with Discussion-Informed Supplementary Material, relative to their baseline understanding with the Original News only, averaged across multiple news articles.}
    \begin{tabular}{lrrrr}
    \toprule
          & \multicolumn{1}{l}{age 6-12} & \multicolumn{1}{l}{age 12-18} & \multicolumn{1}{l}{age 18-35} & \multicolumn{1}{l}{age above 35} \\
    \midrule
    Improve for Vanillar Supplement & 0.0371 & 0.0386 & 0.0408 & 0.0181 \\
    Improve for Discussion Supplement & 0.1381 & 0.0569 & 0.0183 & 0.0423 \\
    \bottomrule
    \end{tabular}%
  \label{tab:Average Improvement in Agent Comprehension for age}%
\end{table*}%

\section{Impact Statement}

This paper explores the application of large language models in the field of communication studies, with a focus on understanding and harnessing the potential of these models to enhance information dissemination and public engagement. The broader impact of our work is significant, as it aims to contribute to the development of more effective and ethical communication strategies in the digital age.

Ethically, we have implemented rigorous protocols to ensure the data used for training our models is as unbiased and representative as possible, and we have developed methods to detect and mitigate the spread of false information.

Looking to the future societal consequences, our work has the potential to influence how educational content is delivered, how news media can personalize information for audiences, and how businesses can engage with customers in a more meaningful way. We recognize the importance of maintaining transparency and accountability in the use of these powerful tools, and we advocate for ongoing dialogue among technologists, ethicists, and policymakers to guide the responsible development and deployment of AI technologies.

\section{Conclusion}
\label{conclusion}
In this work, we addressed the critical challenge of varied news comprehension among diverse audiences, a problem exacerbated by differences in domain expertise and age, which can hinder informed public discourse and civic engagement. We introduced a novel LLM-based agent framework designed to simulate these diverse audience perspectives through an iterative discussion process. This framework uniquely leverages the interactions and questions among agents to diagnose specific comprehension gaps and points of confusion within a news article. Through a series of experiments with agents simulating different occupational experts and age groups, we demonstrated the framework's efficacy in identifying these understanding deficits. Our findings show that supplementary material generated based on the insights from these simulated discussions leads to a statistically significant and consistent improvement in agent comprehension across all tested perspectives, significantly outperforming supplementary material generated by a vanilla LLM without explicit gap identification. Furthermore, our analysis revealed that optimal understanding gains were achieved around the third iteration, providing valuable guidance for computational efficiency. These results underscore the utility and efficiency of our framework in enhancing news comprehension. By providing a systematic method for identifying audience-specific challenges and generating targeted explanations, this work offers a valuable tool for news disseminators to create more accessible content, thereby contributing to a more informed and engaged public capable of navigating the complexities of modern news. 

A key strength and contribution of our framework lies in its extendability and flexibility beyond the specific agent types demonstrated. While we utilized agents simulating distinct occupational domains and age groups as illustrative examples, the framework is designed to be highly customizable. Users can define and input their own specific target audience profiles or roles, articulating the knowledge, perspective, and potential points of confusion characteristic of their intended readers. The framework can then instantiate agents based on these user-defined characteristics, simulate their discussion, identify comprehension gaps specific to those custom profiles, and generate tailored supplementary material accordingly. This capability allows the framework to be adapted and applied across a wide array of specific user needs and audience segments, not limited to our initial examples.

\section{Future Works}
\label{appendix:limitation}
In this paper, we have employed advanced deep learning techniques to develop an LLM for content creation assistance, demonstrating its effectiveness in aiding users in generating informative and inspiring content. However, our study primarily focuses on the English language context, limiting the generalization of our findings to other languages. Future research could extend this work by exploring the application of LLMs in multilingual content creation assistance. Furthermore, while our LLM exhibits promising performance in assisting users, the nuances of human creativity and the potential for unintended biases in AI-generated content remain concerns. Therefore, interdisciplinary collaborations involving experts in AI ethics, communication studies, and linguistics are essential to further refine our LLM, ensuring it not only enhances creativity but also respects cultural and ethical boundaries.


\bibliographystyle{unsrt}  
\bibliography{references}

\begin{thebibliography}{10}

\bibitem{fletcher2017news}
Richard Fletcher and Rasmus~Kleis Nielsen.
\newblock Are news audiences increasingly fragmented? a cross-national comparative analysis of cross-platform news audience fragmentation and duplication.
\newblock {\em Journal of communication}, 67(4):476--498, 2017.

\bibitem{sittar2024news}
Abdul Sittar, Dunja Mladeni{\'c}, and Marko Grobelnik.
\newblock News dissemination: a semantic approach to barrier classification.
\newblock {\em Journal of Intelligent Information Systems}, pages 1--31, 2024.

\bibitem{arndt2025media}
Thomas Arndt.
\newblock {\em Media Bias: Examining the Facts}.
\newblock Bloomsbury Publishing USA, 2025.

\bibitem{yanow2015making}
Dvora Yanow.
\newblock Making sense of policy practices: Interpretation and meaning.
\newblock In {\em Handbook of critical policy studies}, pages 401--421. Edward Elgar Publishing, 2015.

\bibitem{ouyang2025interpreting}
Leyi OuYang.
\newblock Interpreting public sentiment in diplomacy events: A counterfactual analysis framework using large language models.
\newblock {\em Available at SSRN 5286005}, 2025.

\bibitem{mesmer2024glorified}
Kelsey Mesmer, M~Rosie Jahng, Jill Wurm, and Najma Akther.
\newblock “glorified minute takers”: Journalists’(mis) handling of scientific uncertainty during the covid-19 pandemic.
\newblock {\em Science Communication}, 46(6):790--815, 2024.

\bibitem{johnsen2024large}
Maria Johnsen.
\newblock {\em Large language models (LLMs)}.
\newblock Maria Johnsen, 2024.

\bibitem{mo2024large}
Yuhong Mo, Hao Qin, Yushan Dong, Ziyi Zhu, and Zhenglin Li.
\newblock Large language model (llm) ai text generation detection based on transformer deep learning algorithm.
\newblock {\em arXiv preprint arXiv:2405.06652}, 2024.

\bibitem{rachmad2022cognitive}
Yoesoep~Edhie Rachmad.
\newblock Cognitive communication theory, 2022.

\bibitem{wang2023sparsity}
Dan Wang, Zhi Chen, Ionu{\c{t}} Florescu, and Bingyang Wen.
\newblock A sparsity algorithm for finding optimal counterfactual explanations: Application to corporate credit rating.
\newblock {\em Research in International Business and Finance}, 64:101869, 2023.

\bibitem{berger2016social}
Peter Berger and Thomas Luckmann.
\newblock The social construction of reality.
\newblock In {\em Social theory re-wired}, pages 110--122. Routledge, 2016.

\bibitem{tulving1972episodic}
Endel Tulving et~al.
\newblock Episodic and semantic memory.
\newblock {\em Organization of memory}, 1(381-403):1, 1972.

\bibitem{squire2004memory}
Larry~R Squire.
\newblock Memory systems of the brain: a brief history and current perspective.
\newblock {\em Neurobiology of learning and memory}, 82(3):171--177, 2004.

\bibitem{amodio2019social}
David~M Amodio.
\newblock Social cognition 2.0: An interactive memory systems account.
\newblock {\em Trends in Cognitive Sciences}, 23(1):21--33, 2019.

\bibitem{eysenck2020semantic}
Michael~W Eysenck.
\newblock Semantic memory and stored knowledge.
\newblock In {\em Memory}, pages 207--235. Routledge, 2020.

\bibitem{xu2025mem}
Wujiang Xu, Zujie Liang, Kai Mei, Hang Gao, Juntao Tan, and Yongfeng Zhang.
\newblock A-mem: Agentic memory for llm agents.
\newblock {\em arXiv preprint arXiv:2502.12110}, 2025.

\bibitem{fan2023space}
Carina~L Fan, H~Moriah Sokolowski, R~Shayna Rosenbaum, and Brian Levine.
\newblock What about “space” is important for episodic memory?
\newblock {\em Wiley Interdisciplinary Reviews: Cognitive Science}, 14(3):e1645, 2023.

\bibitem{mayes2001theories}
Andrew~R Mayes and Neil Roberts.
\newblock Theories of episodic memory.
\newblock {\em Philosophical Transactions of the Royal Society of London. Series B: Biological Sciences}, 356(1413):1395--1408, 2001.

\bibitem{ewen2021procedural}
Joshua Ewen and Stewart Mostofsky.
\newblock Procedural memory.
\newblock In {\em Encyclopedia of Autism Spectrum Disorders}, pages 3678--3684. Springer, 2021.

\bibitem{stein2019development}
Nancy~L Stein and Christopher~A Miller.
\newblock The development of memory and reasoning skill in argumentative contexts: Evaluating, explaining, and generating evidence.
\newblock In {\em Advances in instructional psychology}, pages 285--335. Routledge, 2019.

\bibitem{kaur2016news}
Gurmeet Kaur and Karan Bajaj.
\newblock News classification and its techniques: a review.
\newblock {\em IOSR Journal of Computer Engineering}, 18(1):22--26, 2016.

\bibitem{mortensen2006human}
C~David Mortensen.
\newblock {\em Human conflict: Disagreement, misunderstanding, and problematic talk}.
\newblock Rowman \& Littlefield, 2006.

\bibitem{cao2024ecc}
Yupeng Cao, Zhi Chen, Qingyun Pei, Nathan Lee, KP~Subbalakshmi, and Papa~Momar Ndiaye.
\newblock Ecc analyzer: Extracting trading signal from earnings conference calls using large language model for stock volatility prediction.
\newblock In {\em Proceedings of the 5th ACM International Conference on AI in Finance}, pages 257--265, 2024.

\bibitem{cao2024risklabs}
Yupeng Cao, Zhi Chen, Qingyun Pei, Fabrizio Dimino, Lorenzo Ausiello, Prashant Kumar, KP~Subbalakshmi, and Papa~Momar Ndiaye.
\newblock Risklabs: Predicting financial risk using large language model based on multi-sources data.
\newblock Technical report, 2024.

\bibitem{cao2024catmemo}
Yupeng Cao, Zhiyuan Yao, Zhi Chen, and Zhiyang Deng.
\newblock Catmemo@ ijcai 2024 finllm challenge: Fine-tuning large language models using data fusion in financial applications.
\newblock In {\em Proceedings of the Eighth Financial Technology and Natural Language Processing and the 1st Agent AI for Scenario Planning}, pages 174--178, 2024.

\bibitem{yu2024finmem}
Yangyang Yu, Haohang Li, Zhi Chen, Yuechen Jiang, Yang Li, Denghui Zhang, Rong Liu, Jordan~W Suchow, and Khaldoun Khashanah.
\newblock Finmem: A performance-enhanced llm trading agent with layered memory and character design.
\newblock In {\em Proceedings of the AAAI Symposium Series}, volume~3, pages 595--597, 2024.

\bibitem{yu2024fincon}
Yangyang Yu, Zhiyuan Yao, Haohang Li, Zhiyang Deng, Yuechen Jiang, Yupeng Cao, Zhi Chen, Jordan Suchow, Zhenyu Cui, Rong Liu, et~al.
\newblock Fincon: A synthesized llm multi-agent system with conceptual verbal reinforcement for enhanced financial decision making.
\newblock {\em Advances in Neural Information Processing Systems}, 37:137010--137045, 2024.

\bibitem{park2023generative}
Joon~Sung Park, Joseph O'Brien, Carrie~Jun Cai, Meredith~Ringel Morris, Percy Liang, and Michael~S Bernstein.
\newblock Generative agents: Interactive simulacra of human behavior.
\newblock In {\em Proceedings of the 36th annual acm symposium on user interface software and technology}, pages 1--22, 2023.

\end{thebibliography}

\end{document}